\documentclass[longbibliography,twocolumn,showpacs,amsmath,amssymb,floatfix]{revtex4-1}

\usepackage{color}
\usepackage{graphicx}

\begin{document}

\bibliographystyle{utcaps}
\title{Pinning the order: the nature of quantum criticality in the Hubbard model on honeycomb lattice}
\author{Fakher F. Assaad$^{1,3}$  and Igor  F. Herbut$^{2,3}$ }
\affiliation{$^{1}$Institut f\"ur Theoretische Physik und Astrophysik,\\
Universit\"at W\"urzburg, Am Hubland, D-97074 W\"urzburg, Germany \\
$^{2}$  Department of Physics, Simon Fraser University, Burnaby, British Columbia, Canada V5A 1S6 \\
$^{3}$ Max-Planck-Institut f\"ur Physik Komplexer Systeme, N\"othnitzer Str. 38, 01187 Dresden, Germany}
\date{\today}

\begin{abstract}
In numerical simulations, spontaneously broken symmetry is often detected by computing  two-point correlation  functions of the appropriate local order parameter. This approach, however, computes the square of the local order parameter, and so when it is {\it small}, very large system sizes at high precisions are required to obtain reliable results.  Alternatively, one can
pin the order by introducing a local symmetry breaking field, and then measure the induced local order parameter infinitely far from the pinning center. The method is tested here at length for the Hubbard model on honeycomb lattice, within the realm of the projective auxiliary field quantum Monte Carlo algorithm.  With our enhanced resolution we find a direct and continuous quantum phase transition between the semi-metallic and the insulating antiferromagnetic states with increase of the interaction. The single particle gap in units of the Hubbard $U$ tracks the staggered magnetization. An excellent data collapse is obtained by finite size scaling, with the values of the critical exponents in accord with the Gross-Neveu universality class of the transition.
\end{abstract}

\maketitle

\section{Introduction}
Detecting spontaneous symmetry broken phases in numerical simulations  often relies on the measure of correlation function.
For instance, the magnetically ordered phase is characterized by long ranged spin-spin correlations,  whereas
the superconducting state exhibits long ranged  pair correlations in the appropriate symmetry channel.  A fundamental caveat
with such an approach is that one measures the {\it square} of the order parameter.  If the later is small, the
quantity one attempts to obtain by extrapolating numerical data to the thermodynamic limit is quadratically smaller. As a
consequence, very large system sizes at high precision are required in addition to an appropriate finite size extrapolation formula.
	
The aim of  this article is  two-fold.  We will  document a simple and very efficient  alternative method  to detect  magnetically ordered phases in SU(2) invariant Hubbard type models in the realm of projective  quantum Monte Carlo methods. With the  enhanced resolution,  we will revisit the semi-metal to insulator transition on graphene's honeycomb lattice, which has recently been under considerable debate \cite{Meng10,Sorella12}.

Honeycomb lattice is a bipartite, non-frustrated lattice, which at half filling and small Hubbard repulsion $U$ hosts the semi-metallic state of electrons, as in graphene.  When the repulsion is increased one expects eventually a phase transition into an insulating state with antiferromagnetic order \cite{Sorella92,Paiva05,Herbut06}, which due to gapless Dirac fermionic excitations being present on the semimetallic side, should belong to a particular, Gross-Neveu universality class \cite{Herbut06,Herbut09a}.

Starting from the strong coupling limit and  noting that the insulator to metal transition occurs at values of the Hubbard interaction lesser than the bandwidth, allows for the proliferation of higher order ring exchange terms  in an effective spin model aimed at describing the magnetic insulating state in the vicinity of the  transition \cite{Schmidt12}.   This point of view opens the possibility that the melting of the magnetic order is unrelated to the the  metal-insulator transition.   Recent quantum Monte Carlo calculations \cite{Meng10} suggested that there is an intermediate spin liquid phase with a single-particle gap  but no magnetic ordering, separating the semi-metal and  magnetic insulator.  Similar results have been put forward  for the related $\pi$-flux model on the square lattice \cite{Chen2012}.  The results of Ref. \cite{Meng10}  have been challenged  by recent studies.    Entropy calculations  do not favor  ground state degeneracy as expected  for the $Z_2$ spin liquid \cite{Clark2013}.   Moreover,  Ref.  \cite{Sorella12}
shows that extrapolating from  significantly larger system sizes would suggest almost complete disappearance of the spin-liquid from the phase diagram.  The latter conclusion is reinforced here, where we find excellent data collapse and {\it identical} finite-size scaling of both single-particle gap and staggered magnetization, with the distinct values of critical exponents,  in accord with the Gross-Neveu universality class \cite{Herbut06,Herbut09a}.

 Form the technical point of view, our approach is very similar in spirit to  an approach considered in Ref. \onlinecite{White07}.  By introducing  a local magnetic field at say the origin,  we explicitly break  the  SU(2) spin symmetry.  In  the presence of long range order and  in the thermodynamic limit,  any field will pin the order   along the direction of the external field.   Thereby,  order can be detected by computing directly the magnetization infinitely far from  the pinning field.  The upside of such an approach is that one measures directly the order parameter rather than its square.  This amounts to evaluating a single particle quantity  which is often much more stable than correlation functions.   The downsides are three-fold.   One explicitly breaks SU(2) spin symmetry such that  spin sectors mix and it becomes computationally  more expensive to reach the ground state.  Since the computational cost scales linearly with the projection parameter,  this problem is tractable.  The second  difficulty lies in the ordering of limits.
To obtain results which are independent on the magnitude of the pinning field,  it is important to first take the thermodynamic limit and then the limit of infinite distance from the pinning field. In a practical implementation, this ordering of limits has as a consequence some leftover dependence of the magnetization  on the magnitude of the pinning field.  This is  particularly  visible when the pinning field is {\it small}.  The final drawback is that  it is not always possible to introduce a  pinning   field without  generating a negative sign problem. For instance, in the Kane-Mele Hubbard model \cite{Hohenadler10,Hohenadler12,Rachel10}, the  spin order lies in the x-y plane. Adding a magnetic field along this quantization axis introduces a sign problem. On the other hand, the method is applicable to SU(N) symmetric Hubbard-Heisenberg models \cite{Assaad04}.

The organization and main results  of the article are the following.  We will focus on the Hubbard model on honeycomb lattice at the filling one half, for which the presence of an intermediate spin-liquid phase has been controversial \cite{Meng10,Sorella12}. After introducing and testing the approach in the next section, we will provide a phase diagram of the Hubbard model in Sec. \ref{Sec:Phase}. The data points to the fact that the staggered moment follows rather precisely the single particle gap, when the latter is measured in the natural units of the Hubbard $U$,
suggesting a direct quantum phase transition between the semi-metallic and the insulating antiferromagnetic phases. Furthermore, an excellent finite size scaling of the data for both the staggered magnetization and the single particle gap is found by assuming the values of the critical exponents $\beta= 0.79$ and $\nu=0.88$. These are the values found in the first order expansion for the Gross-Neveu-Yukawa field theory of this quantum phase transition \cite{Herbut09a}, around its upper critical (spatial) dimension of three.  Altogether, the data strongly supports the existence of a single quantum critical point separating the semi-metallic and the insulating antiferromagnetic phases of the Hubbard model, with the quantum criticality  belonging to  the Gross-Neveu universality class \cite{Herbut06}.

\section{Model and method}
As in Ref. \cite{Meng10},  we will consider the half-filled  Hubbard model on the Honeycomb lattice

\begin{equation}
  H_{tU} =  -t\sum_{\langle \bf{i},\bf{j} \rangle, \sigma}   c^{\dagger}_{\bf{i}, \sigma} c^{}_{\bf{j}, \sigma}      + U \sum_{\bf{i}}
\left( n_{\bf{i},\uparrow} - 1/2\right)  \left( n_{\bf{i},\downarrow} - 1/2 \right).
\end{equation}
The hopping is restricted to nearest  neighbors  so that the bipartite nature of the lattice allows us to avoid the negative sign problem.

 Generically, to detect anti-ferromagnetic ordering  we compute spin-spin correlations:
\begin{equation}
\label{Correl}
m =    \lim_{L\rightarrow  \infty }\sqrt{  \frac{1}{N}  \sum_{ {\bf{i}}=1}^{N}  e^{i {\mathbf{Q}} \cdot {\bf{i}} } \langle {\bf{S}}_{\bf{0}} \cdot {\bf{S}}_{\bf{i}} \rangle_{H_{tU}} }.
\end{equation}
Here $ N = 2 L^2$ corresponds to the  number of orbitals, and $L$ is the linear length of the lattice.
A finite value of  $m$ signalizes long range order  and is equivalent to spontaneous symmetry breaking.  In particular,  including  a magnetic field term  with  appropriate  Fourier component,
\begin{equation}
  H_h  =    h \sum_{\bf{i}}  e^{i \mathbf{Q}\cdot \bf{i}}  S^{z}_{i} ,
\end{equation}
gives
\begin{equation}
	m
=    \lim_{h \rightarrow 0} \lim_{L\rightarrow \infty}    \frac{1}{L^2} \sum
_{\mathbf{i}} e^{i  \bf{Q} \cdot \bf{i}}\langle S^{z}_{\bf {i}} \rangle
_{H_{tU} +  H_h}.
\end{equation}
The ordering of limits is crucial.  One first has to take the  thermodynamic limit to allow for the   collapse of Anderson's tower of states, and then the  limit
 of vanishing magnetic field $h$. Such an approach was for instant used in Ref. \onlinecite{Ulybyshev2013}.

\begin{figure}
\includegraphics[width=\linewidth]{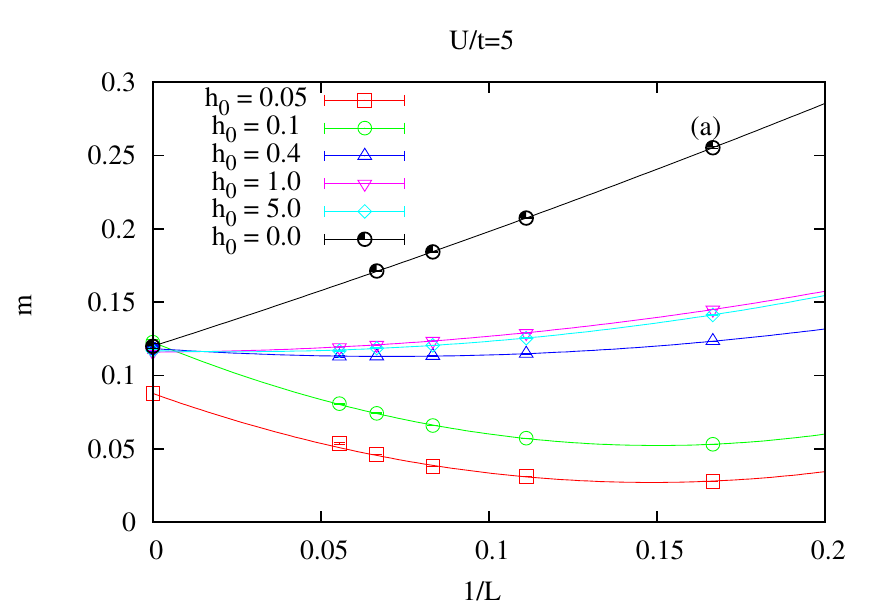}
\includegraphics[width=\linewidth]{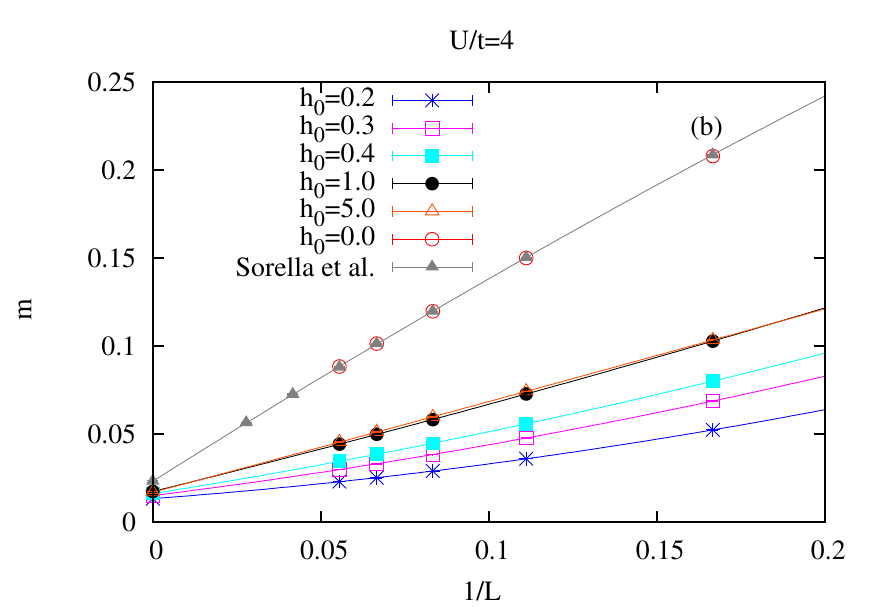}
\caption{ Comparison of the  pinning field  and correlation function approaches to determine  the staggered moment at U/t=5 (a) and U/t = 4 (b). The data sets at $h_0 = 0$ corresponds to the  correlations functions, and values of $\theta t = 40$ are sufficient to  converge to the ground state.  For non-vanishing  pinning  fields  projection parameters
$\theta t = 320 $ are required to guarantee convergence.   We have used $\Delta \tau t = 0.1$ which for  the symmetric Trotter decomposition yields converged results within our numerical accuracy.
At $U/t = 4$ comparison with results of Ref. \onlinecite{Sorella12} shows excellent agreement.  Lines corresponds to
least square fits to the form $ a + b/L + c/L^2 $.}

\label{fig:Comparison}
\end{figure}

It is more convenient to consider a local field  since,  as we will see below, this lifts  the burden of  taking   the limit   $h \rightarrow 0$ numerically.
The local pinning field is given by the term
\begin{equation}
H_{loc}  = h_0  S^{z}_{\bf{0}}
\end{equation}
in the Hamiltonian.
Using  the representation $ \delta_{\bf{i},\bf{0}} = \frac{1}{L^2} \sum_{\bf{q}}  e^{i \bf{q}\cdot \bf{i}} $ of the Kronecker symbol shows that each Fourier component comes with an amplitude  $h_0/L^2$ so  that  taking the thermodynamic limit is equivalent to taking the amplitude of the relevant  Fourier component to zero.
With  the local field construction, the appropriate ordering  of limits  for an $L \times L$ lattice reads:
\begin{equation}
\label{Lim.eq}
m = \lim_{\bf{i}\rightarrow \infty } \lim_{L \rightarrow \infty}   e^{i \bf{Q}\cdot \bf{i} }\langle S^{z}_{\bf{i}} \rangle_{H_{tU} + H_{loc} }.
\end{equation}

That is,  one first has to take the  thermodynamic limit  -- again to guarantee the collapse of the tower of states   in the presence of  long range order -- and only then can one take the distance from the pinning center to infinity
\footnote{Clearly, one cannot exchange the limits in Eq. \ref{Lim.eq}.  However,  one  can measure the magnetization at the largest distance  available on the lattice, thereby tying  together the thermodynamic and infinite distance from the pinning center limits.  This procedure can lead to spurious results.  }.
In other words, the distance from the pinning center sets an energy scale which has to be larger than the finite size  spin gap.
As an efficient estimator  for the evaluation of the  ordered moment we thus propose:
\begin{equation}
\label{Mag}
      m =   \lim_{L\rightarrow \infty}  \frac{1}{L^2}   \sum_{\bf{i}}    e^{i \bf{Q}\cdot \bf{i} }\langle S^{z}_{\bf{i}} \rangle_{H_{tU} + H_{loc} }.
\end{equation}

We have tested the above approach  for the Hubbard model on the honeycomb lattice.   Ground state calculations were carried out with the projective auxiliary field   Quantum Monte Carlo (QMC)  algorithm which is based on the equation:
\begin{equation}
	\langle  O \rangle_{H}   = \lim_{\theta \rightarrow \infty}
    \frac{  \langle \Psi_{T} | e^{- \theta H/2} O e^{- \theta H/2}  |\Psi_{T} \rangle  }
	   { \langle \Psi_{T} | e^{- \theta H}  |\Psi_{T} \rangle }.
\end{equation}
Here $\theta$ is a projection parameter, and the trial wave function is required to be  non-orthogonal to the ground state.
For  $H = H_{tU}   + H_{loc}$  the  inclusion of the magnetic field does not generate a negative sign problem.     We have chosen the  trial wave
function to be the ground state of the  non-interacting Hamiltonian $H_T = H_{t} + H_{loc} $   in the $S^{z} = 0 $ sector.   The implementation of the algorithm follows closely  Refs. \onlinecite{Meng10,Assaad08_rev}.  The major difference is the use of a symmetric Trotter
breakup which ensures hermiticity  of the imaginary time propagator for  any value of the  time discretization $\Delta \tau$. It also  leads to  smaller systematic errors.

Figure \ref{fig:Comparison} (a) plots the local moment   at $U/t=5 $ using different methods. In this case, magnetic ordering  is robust such  that various  approaches can be compared.  The data set  at $h
_0 = 0$  corresponds to the correlation functions of Eq. \ref{Correl}.  For this set of runs we used a spin-singlet trial wave function and the projection parameter $\theta t = 40$ suffices to guarantee convergence to the ground state.  This quick convergence  stems from the fact that  the  trial wave function is orthogonal  to the low lying spin excitations \cite{Capponi00}.
The runs at finite values of the pinning field correspond to the quantity of  Eq. \ref{Mag}.
 In the presence of a finite pinning field, SU(2) spin symmetry is broken and the trial wave function overlaps with all spin sectors.  Consequently,  a large value of the projection parameter $\theta t = 320$ is required to guarantee convergence to the ground state within the quoted  accuracy.    Note that the  CPU time scales only linearly with the projection parameter, so that such large projection parameters are still numerically tractable.  It is also worth pointing out that  the observable of Eq. \ref{Mag} corresponds to  a single particle quantity and shows very little fluctuations.
 As evident in Fig. \ref{fig:Comparison}(a), finite size effects are strongly dependent on the specific choice of the pinning field. Nevertheless, convergence to values consistent with  the generic approach based on Eq. \ref{Correl} is obtained for relatively {\it large}  values of the pinning field.  If the pinning field is chosen too  {\it small}, larger lattices are apparently required to  insure that the finite size spin gap,  set by $v/L$ with $v$ the spin wave velocity,  is smaller than the energy scale set by the pinning field.  This expectation is confirmed by the data, which shows a systematic upturn as a function of the system size for smaller values of the  pinning field.  Such a non-monotonic finite size behavior complicates a finite size scaling analysis. For this reason, we propose to use a relatively large value of the pinning field
\footnote{Here large means comparable to the bandwidth. If the pinning field  is much larger than the bandwidth, charge fluctuations on the pinning site will be blocked by an energy scale set by $h_0$.   Thereby in the limit $h_0 \rightarrow \infty$  the pinning site effectively drops out of the Hamiltonian and no symmetry breaking occurs. The very slight drop in the magnetization at $h_0 =5$ and on small lattices in  Fig. \ref{fig:Comparison}(a) could be a precursor of this effect. }.
at U/t=5 (a) and U/t = 4 (b).
At U/t = 4, the local moment is  smaller,  and hard to detect.   The  $h_0 = 0$ data set of  Fig. \ref{fig:Comparison}(b)   compares our results for the correlation function of Eq. \ref{Correl} to those of Ref. \onlinecite{Sorella12}.  As apparent,   the agreement  up  to our largest lattice size, $L=18$, is remarkable.  Without  the largest lattice sizes,  $L=24$ and $L=36$,  extrapolation to the thermodynamic limit is hard due to the downward turn  present in the finite size results. The  data sets stemming from the pinning field approach provide an alternative perspective,  and, on the whole, confirm the result of Ref. \onlinecite{Sorella12}.  For the  considered field range  there is considerable scatter in the finite size results, but nevertheless   the extrapolation to the thermodynamic limit seems to be field independent, as expected from the above considerations.

We conclude this section by mentioning that we have tested  the approach for the non-interacting case and  the method successfully demonstrates the absence of long range  magnetic order. Hence  both for critical states as well as for magnetically ordered phases the pinning field approach does provide an efficient tool.  Further testing of this approach for  Heisenberg bilayers is presently under progress \footnote{S. Wessel private communication} .

\section{Phase diagram of the Hubbard model on honeycomb lattice.}
\label{Sec:Phase}
We have used the above approach to revisit the magnetic phase diagram of the Hubbard model on the  honeycomb lattice.
At weak couplings the model is known to have a stable semi-metallic state.  In the strong coupling limit, and due to the  absence of  frustration, an anti-ferromagnetic Mott insulator is present. The nature of the transition between these two states has been studied in the past, \cite{Sorella92, Paiva05, Herbut06} and is presently controversial \cite{Meng10,Sorella12}.

\begin{figure}
\includegraphics[width=\linewidth]{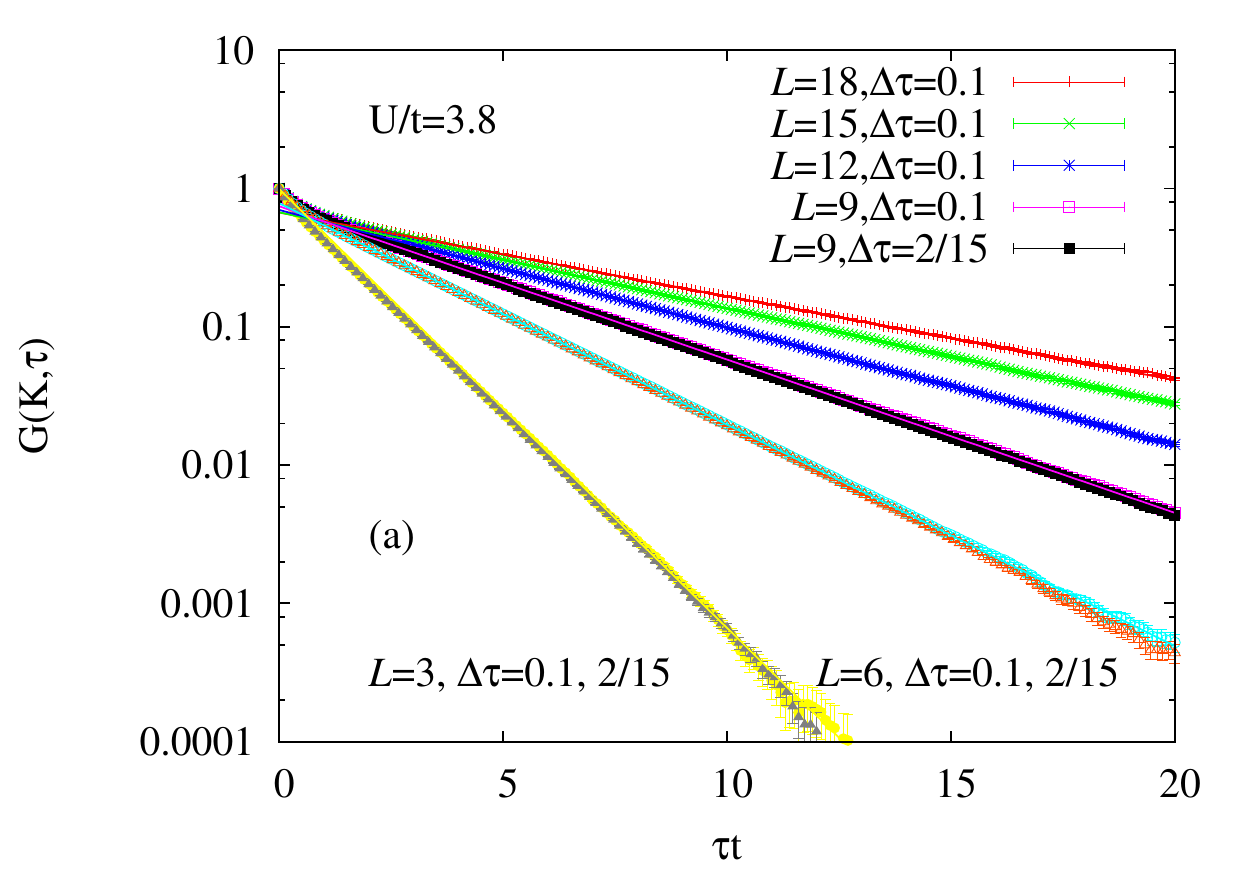} \\
\includegraphics[width=\linewidth]{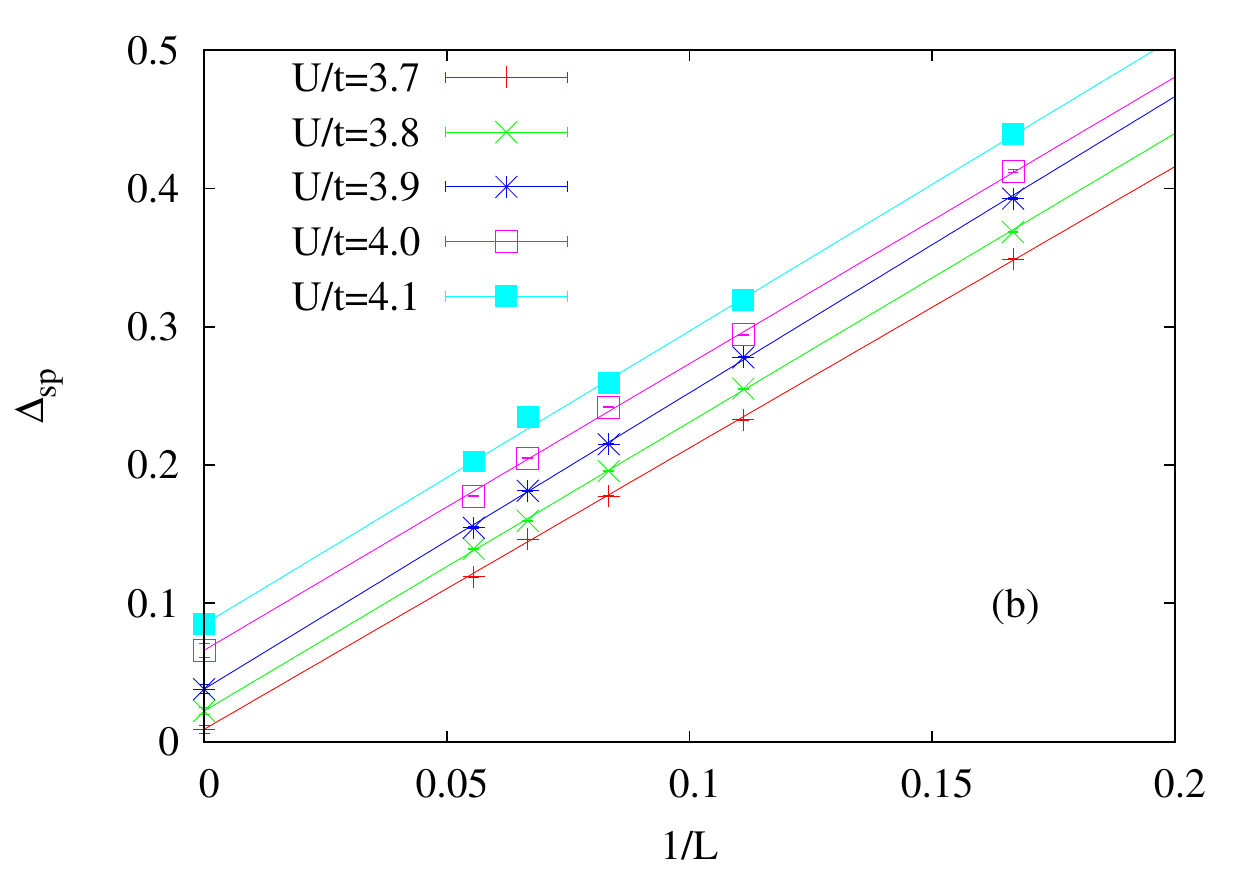} \\
\caption{ Single particle gap. (a) The raw data at $U/t = 3.8$. As apparent, the systematic error stemming from the finite Trotter step is negligible within our accuracy.  The data supports a {\it large }  imaginary time range  consistent with a single exponential decay. Lines are least square fits of the tail of the imaginary time Green function to the form
$ Z e^{-\Delta_{sp} \tau} $. Here $Z$ corresponds to the single particle residue and $\Delta_{sp}$ to the single particle gap.  (b) Size dependence and extrapolation of the single
particle gap. }
\label{fig:Single_particle_gap}
\end{figure}

\subsection{Single particle gap.}
To pin down the  coupling strength beyond which the single particle gap opens, we have  repeated calculations  for  the time displaced single particle  imaginary time Green function at the nodal point:

$G (K,\tau)  =   \sum_{\sigma}\langle c^{\dagger}_{\bf{K},\sigma}(\tau)  c^{}_ {\bf{K},\sigma}(\tau=0)\rangle $.
As evident in Fig. \ref{fig:Single_particle_gap}, and with the symmetric Trotter decomposition,
the Trotter systematic error is negligible  within our accuracy.
Fitting the data to an exponential form allows us to extract the single particle gap,  which we plot as a function of system size in Fig.  \ref{fig:Single_particle_gap}. Assuming a polynomial form for the extrapolation to the thermodynamic limit, we find a small but finite single particle gap for  $U/t  \geq  3.7$. This finding is particularly interesting when compared to the results of  Sorella et al \cite{Sorella12}   which at $U/t =3.8$  point to the absence of  long range magnetic order. This could be taken as an indication for a possible intermediate phase. However, the analysis that we present below suggests a different interpretation.

\subsection{Magnetization from pinning fields.}
\begin{figure}
\includegraphics[width
=\linewidth]{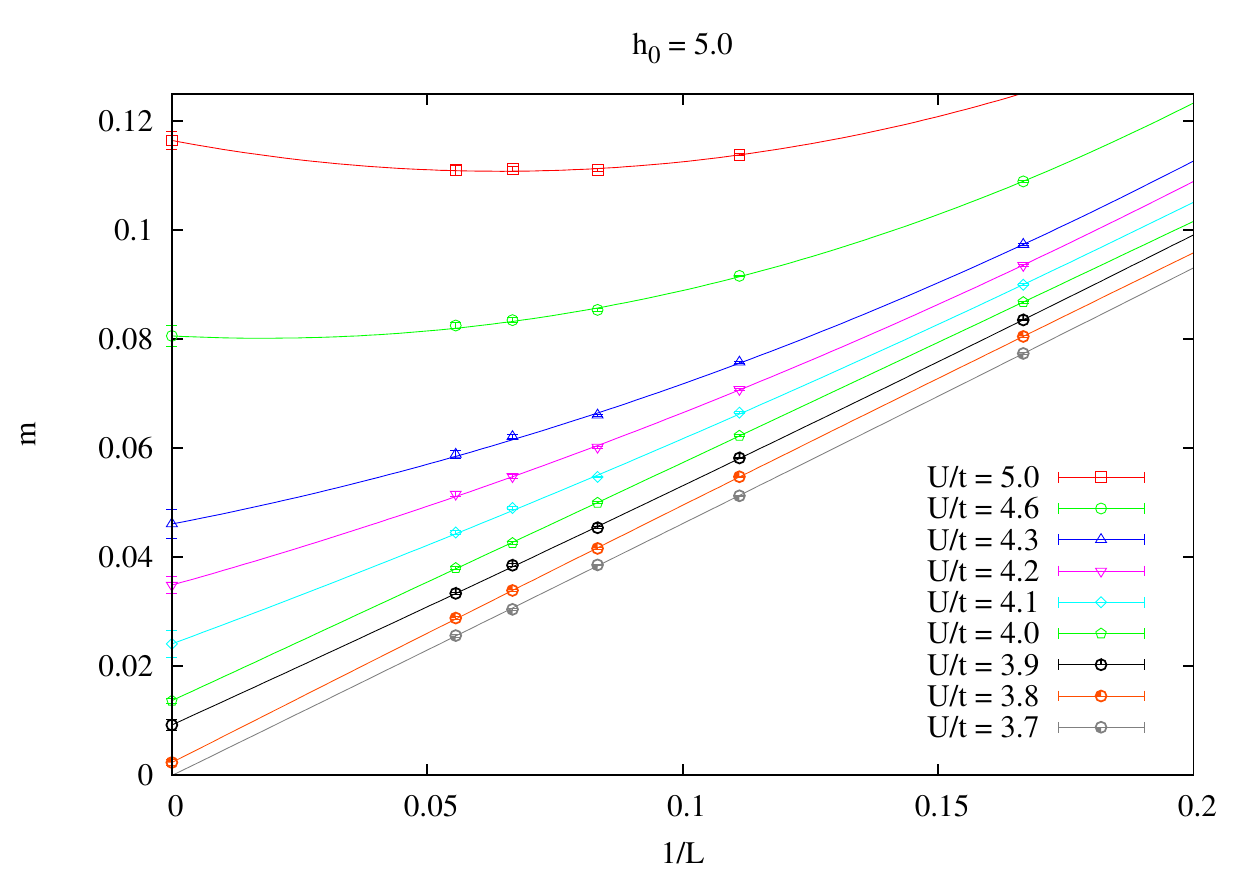} \\
\caption{ Magnetic moment at pinning field $h_0 = 5t$ as a function of $U/t$. Here we have used $\theta t = 320$ and $\Delta \tau t = 0.1$. The solid lines are
least square fits to the form $ a + b/L + c/L^2$. }
\label{Fig:MagU}
\end{figure}

\begin{figure}
\includegraphics[width=\linewidth]{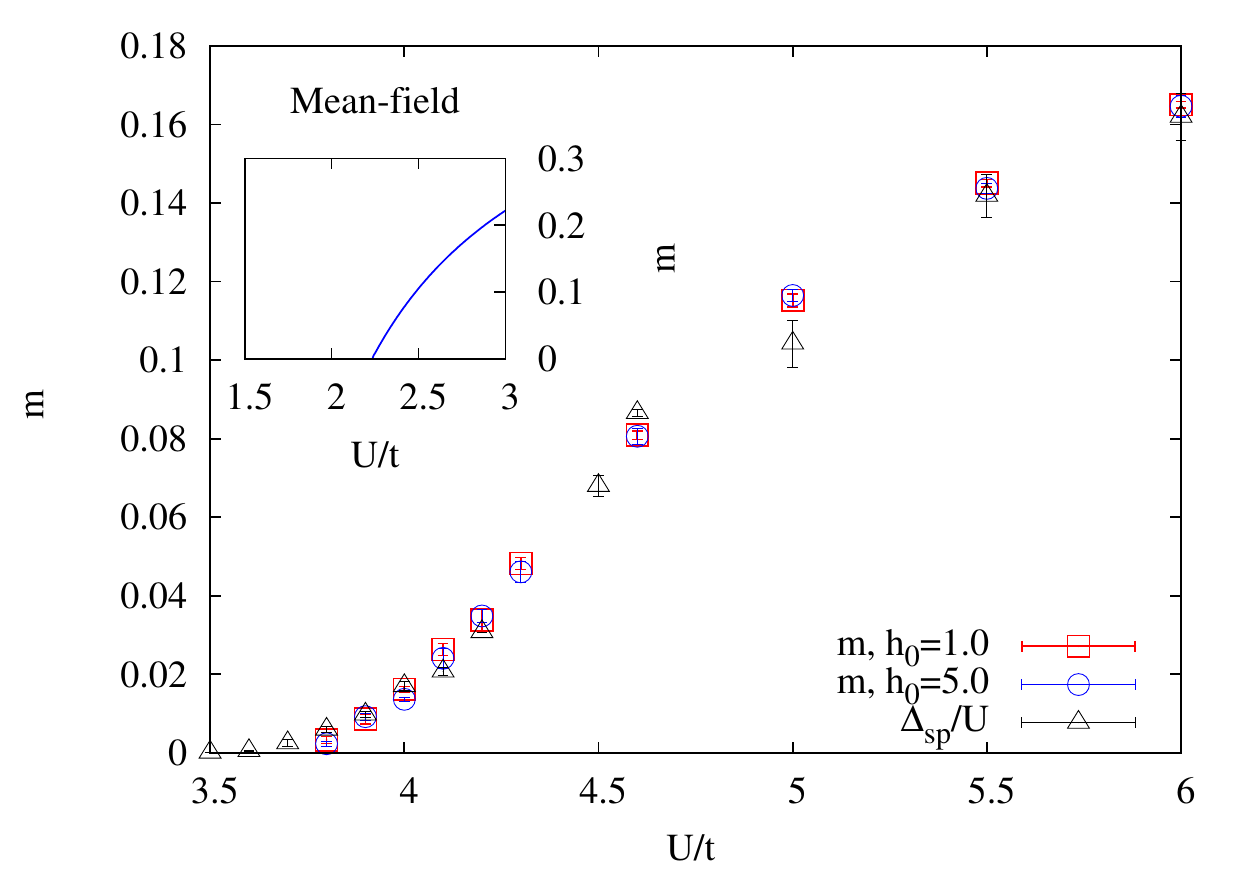} \\

\caption{Staggered moment extrapolated to the thermodynamic limit (see Fig. \ref{Fig:MagU}) for two values of the pinning field. We have equally plotted the single particle gap  in units of $U$. The inset plots the staggered magnetization as obtained from a  mean-field spin density wave Ansatz. }

\label{Fig:Sum}
\end{figure}

We have used the pinning field approach to  compute the  staggered moment as a function of  $U/t$.  The results  at $h_0
= 5 t $ are  reported in Fig.  \ref{Fig:MagU}. The extrapolation to the thermodynamic limit is carried out using a polynomial   scaling up to second order in $1/L$.  Fig. \ref{Fig:Sum} plots the so obtained staggered moment for two choices of
the pinning field as well as the single particle gap.    Several comments are in order.
\begin{itemize}
\item  Within our accuracy, and maybe most importantly, with the polynomial fit used in extrapolating the data to the thermodynamic limit, it appears that the single particle gap  opens  right when magnetic ordering sets in.  The only mismatch is at $U/t=3.7$ where we do  not detect magnetic ordering but we do detect a  small single particle gap.
\item

The QMC data in Fig. \ref{Fig:Sum} shows that  over a wide parameter range, the single particle gap measured in units of the Hubbard U, tracks the  staggered magnetization.
We take this as a strong indication, that the magnetization provides the only relevant scale in the problem, determining directly the single particle gap. We will see below, that this conclusion, based here on a simple, polynomial extrapolation of the finite size data, is also obtained, if a more refined data analysis is performed.

\item   The data in Fig. \ref{Fig:Sum} exhibits an unusual inflection point at approximately $U/t = 4.1$. Such an inflection point is clearly  absent at the mean-field level (see inset of Fig. \ref{Fig:Sum}). We will discuss the implications  of this inflection point in the next section.
Let us finally note, that  in previous calculations \cite{Meng10}  we were unable to resolve  staggered moments  lesser than $ m \simeq 0.03$. We thereby missed this inflection point in the polynomially extrapolated magnetization curve  and concluded the  presence of an intermediate phase
\footnote{In Ref. \cite{Meng10}  a small but finite spin gap with maximal value at $U/t=4$ was reported.  The spin gap is determined  with the  same method as  the single particle gap, by fitting the tail of the  imaginary time displaced spin-spin  correlation functions to a single exponential.   Enhancing   the imaginary time range for the fit produces a slight decrease of the spin gap for the larger lattices sizes. This in turn renders the extrapolation to the thermodynamic limit inconclusive.  }.

\end{itemize}

\subsection{Finite size scaling}
\begin{figure}
\includegraphics[width=\linewidth]{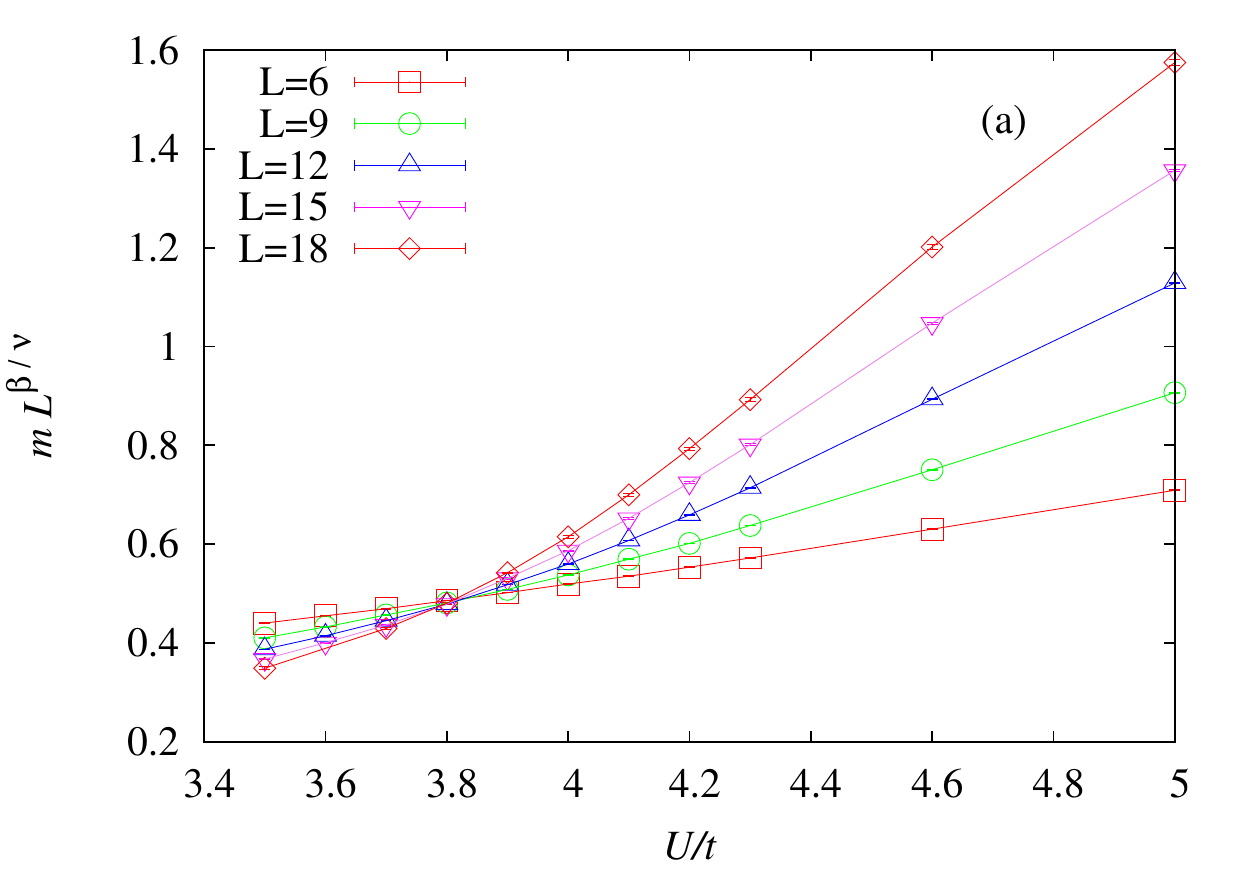} \\

\includegraphics[width=\linewidth]{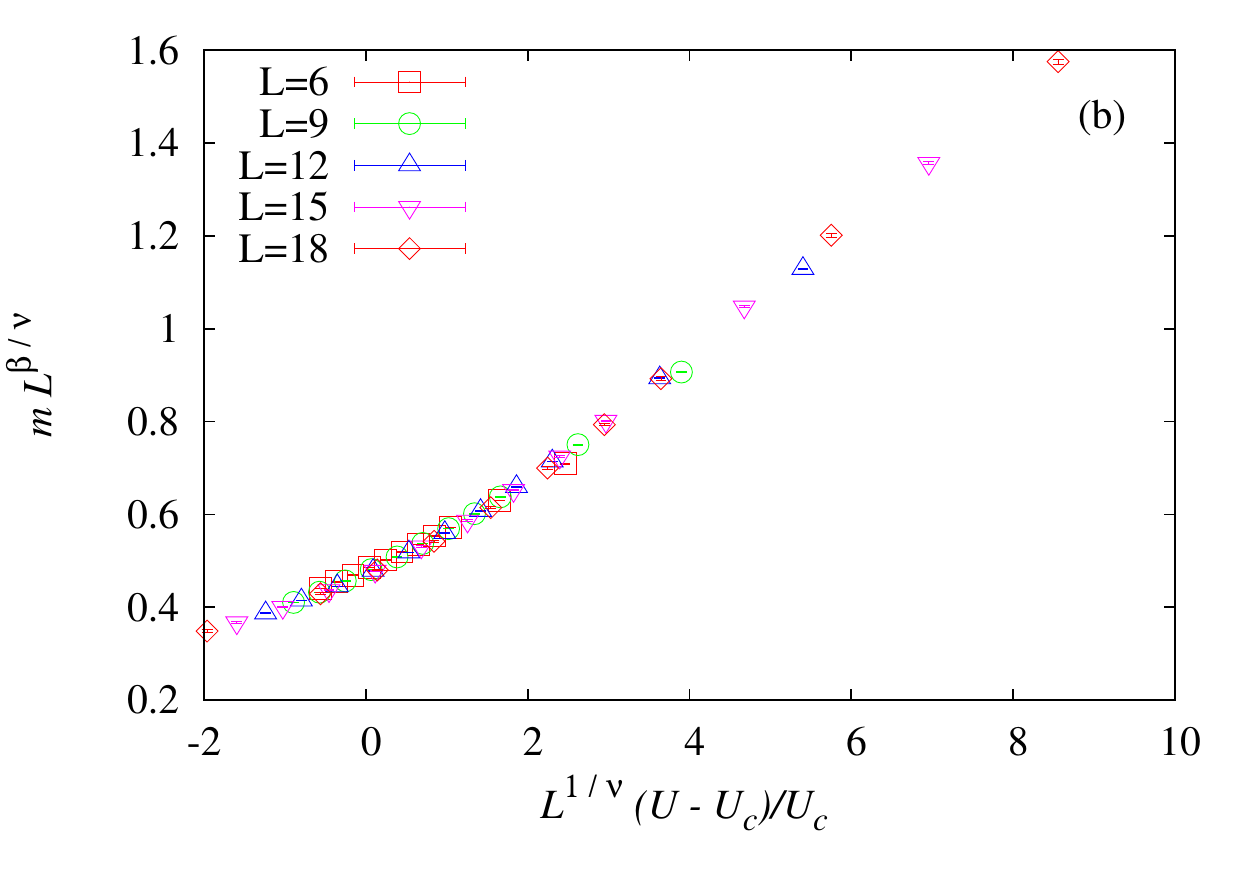}
\caption{  Data collapse  for the magnetization  presented in Fig. \ref{Fig:MagU}.   The exponents are taken for the $\epsilon$-expansion of Ref. \onlinecite{Herbut09a}.
(a)  The crossing point pins down the value of $U_c$.  (b)  The data collapse, using $U_c/t =3.78$.}
\label{Fig:Collapse}
\end{figure}
\begin{figure}
\includegraphics[width=\linewidth]{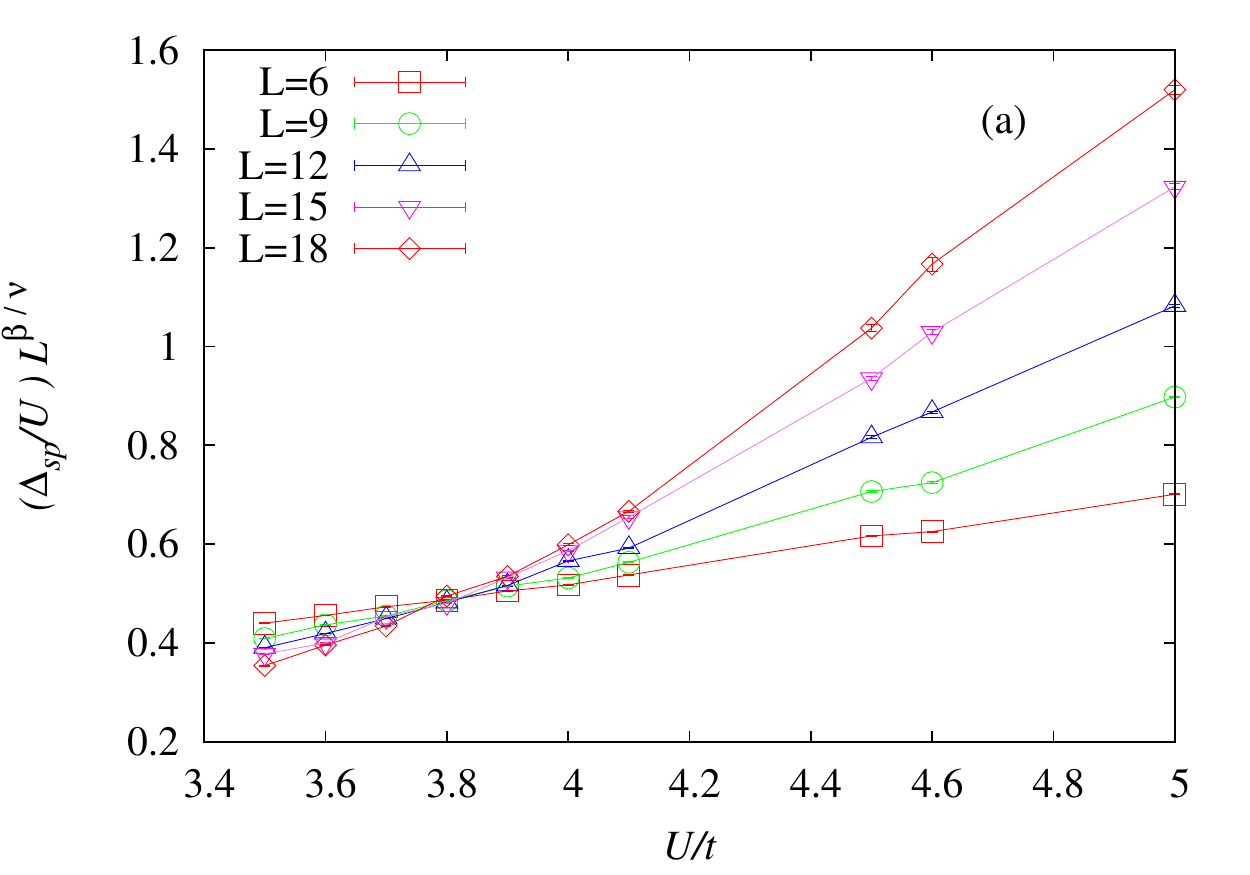} \\
\includegraphics[width=\linewidth]{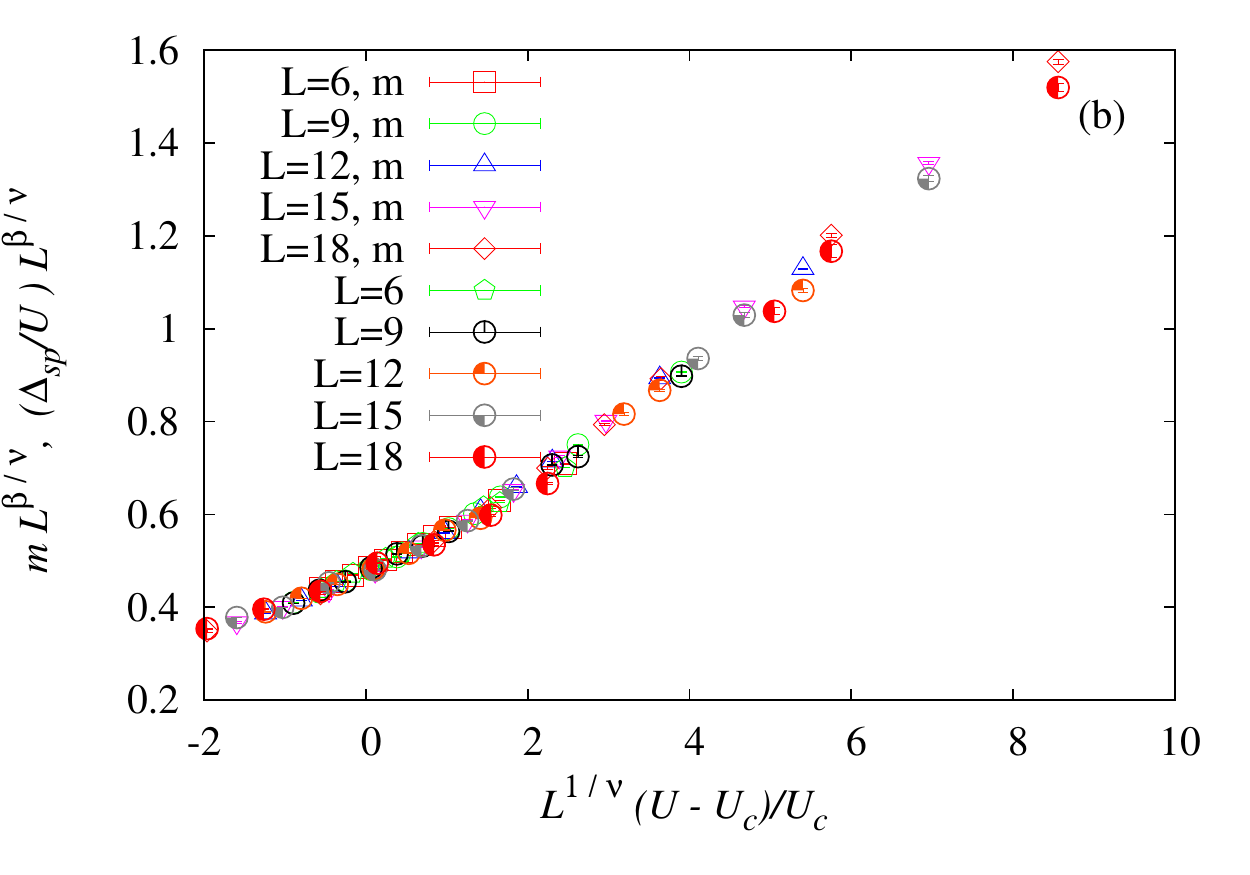}
\caption{  Data collapse  for the single particle gap.   The exponents are taken for the $\epsilon$-expansion of Ref. \onlinecite{Herbut09a}.
(a)  The crossing point pins down the value of $U_c$.  (b)  The data collapse again using $U_c/t =3.78$. For comparison we have included the data for the magnetization.}
\label{Fig:Collapse_QP}
\end{figure}
As mentioned above, one of the particularities of the data presented in Fig. \ref{Fig:Sum}  is the  occurrence of an inflection point at $U/t = 4.1$.  It is a natural question  to ask if this rather peculiar feature may be an artifact of
using a simple polynomial fitting procedure, which one would indeed expect to fail close to criticality. This could result in an overestimation of the magnetization in the vicinity of the critical point between the semi-metallic and the insulating phase of the Hubbard model. As we explain next,
arguments in favor of this conjecture are provided by the large-N treatment of the Gross-Neveu model \cite{Herbut06},  and the $\epsilon$-expansion around three spatial  dimensions  in the equivalent Gross-Neveu-Yukawa field theory, formulated in Ref. \onlinecite{Herbut09a}. Given the order parameter exponent, $\beta$, as well as the correlation length exponent, $\nu$,  the staggered magnetization scales as
\begin{equation}
	  m   \simeq    | U - U_c |^{\beta}  \simeq  \xi^{-\beta/\nu}.
\end{equation}
Using the standard scaling laws \cite{HerbutBook},  the exponent $\beta/\nu$ may conveniently be expressed in terms of the anomalous dimension for the order parameter $\eta$, as
\begin{equation}
   \frac{\beta}{\nu}    \equiv  \frac{1}{2}   \left(  [d+z]  - 2  + \eta  \right),
\end{equation}
where $d+z$ is the effective dimensionality of the system.
If we assume that the Lorentz  invariance is emergent at the critical point, as it indeed is close to the upper critical dimension $d_{up}=3$ of the Gross-Neveu-Yukawa theory \cite{Herbut09a}, and maybe even more generally \cite{Herbut09}, the dynamical critical exponent is $z=1$. If then the anomalous dimension of the order parameter is such that $\eta < 3-d$, we find that the combination of the exponents $\beta/ \nu < 1 $, and  our  polynomial fitting procedure  in the previous section could very well overestimate the value of the staggered magnetization. In fact both the large-N approach \cite{Herbut06} and the expansion around the upper critical dimension \cite{Herbut09a} suggest that this is indeed the case.   Within the first order of the expansion in the parameter
$\epsilon=3-d$,  for example, $\eta = 4\epsilon/5 $, so that
\begin{equation}
 \frac{\beta}{\nu}    = 1  -   \frac{\epsilon}{10}  + {\cal O}\left(  \epsilon^2 \right).
\end{equation}
In two dimensions then,  $\beta/\nu \simeq 0.9 $.

To look for the signs of the Gross-Neveu criticality in the Hubbard model
we have carried out a finite size scaling analysis based on the usual scaling form
\begin{equation}
	 m = L^{-\beta/\nu}   F( L^{1/\nu} (U - U_c) ).
\end{equation}
Figure \ref{Fig:Collapse}(a)  plots $m L^{\beta / \nu} $  versus $U$ for the magnetization data at the fixed field $h_0 = 5$.  (We will omit at this point the second scaling variable, $h_0 L^{y-d}$, since the scaling dimension $y-d= (\epsilon-\eta)/2 \ll 1$.  This second argument of the scaling function, present in principle, is therefore effectively constant at a fixed $h_0$, and its inclusion does not visibly affect the quality of scaling. For further discussion of this point, see the Appendix.) As a guide, we have used the first-order  $\epsilon$-expansion value of $\beta / \nu = 0.9$. Five  curves then all cross at a single point $U_c/t  \simeq 3.8 $,  thereby providing a first non-trivial indication of the critical point. This value of $U_c$  is slightly larger than that obtained with the  polynomial fit. The $\epsilon$-expansion value of the correlation length exponent reads \cite{Herbut09a},
\begin{equation}
	\nu = \frac{1}{2} +  \frac{21}{55} \epsilon +  {\cal O}\left(  \epsilon^2 \right).
\end{equation}
With this  value of $\nu$,  again at $\epsilon = 1$, we obtain an excellent data collapse,  as shown in Fig. \ref{Fig:Collapse}(b).

In accord with the Gross-Neveu-Yukawa  theory,  the numerical data of  Fig.  \ref{Fig:Sum}  support  the interpretation that the magnetization is the only scale in the problem.   To further check this interpretation,  we  have scaled the single single particle gap to the form:
\begin{equation}
	\frac{\Delta_{sp}}{U}  = L^{-\beta/\nu}   \tilde{F}( L^{1/\nu} (U - U_c) )
\end{equation}
again using the $\epsilon$-expansion exponents in two dimensions.
Fig. \ref{Fig:Collapse_QP}(a)   shows  that  the  crossing point of the $ \frac{\Delta_{sp}}{U}  L^{\beta/\nu} $ curves again occur at  $U_c/t \simeq 3.8$ and  Fig. \ref{Fig:Collapse_QP}(b)   shows  the collapse. It is quite remarkable, that within our  precision, the two scaling functions are equal,   $\tilde{F} = F$.

Hence,  the scaling analysis of our  QMC  within the Gross-Neveu scenario is consistent with a single continuous quantum phase transition between the semimetal and the antiferromagnetic insulator, and suggests that the first-order  expansion around the upper critical dimension in the Gross-Neveu-Yukawa  theory may already yield rather accurate values of the critical exponents.

\section{Discussion and conclusions}

We have introduced an alternative method to compute  the staggered magnetization.  By introducing a local magnetic field we  pin the quantization axis of the ordered state. The staggered magnetic moment then corresponds to the  local magnetization infinitely far away from the pinning center. The approach has the major advantage that we compute directly the magnetization  as opposed to its square when measuring correlation functions.   We can therefore expect improved resolution  when the local moment is {\it small}.   One advantage  of the approach is an internal cross-check which requires the  staggered moment to be independent on the numerical value of the pinning field.  We have been able to reach this  internal cross-check   only in the case of {\it large}  pinning fields.   This is consistent with the approach proposed by \cite{White07} where the pinning field is set to infinity on the boundary of the lattice.  If the pinning field is too small, the approach suffers from large and non-monotonic size effects, since the energy scale set by the local magnetic field is unable to overcome the finite size spin gap. Under these circumstances the extrapolated value of the magnetization has the tendency of underestimating the order.    In this article  we have only considered a very specific form of the pinning field.  The number of different  choices of pinning fields  provides a playground for optimization of the approach, and for minimization of the size effects.

The application to the Hubbard model on the honeycomb lattice sheds new light on the phase diagram of this well known problem.
The enhanced precision in comparison to Refs. \onlinecite{Meng10}  and \onlinecite{Sorella12} reveals  that the staggered moment has the same functional form as the single particle gap (as measured in units of $U$). Remarkably,  an excellent data collapse onto a {\it single universal curve} is found in the finite size scaling of both quantities, with the values of the critical exponents characteristic of the Gross-Neveu criticality between the semimetallic and the magnetic insulating phases.

\acknowledgments
 FFA is very indebted to T. Lang,  Z.Y. Meng,  A. Muramatsu and  S. Wessel  for many valuable  comments.   We equally acknowledge discussions with A. Chernyshev, L. Fritz,  M. Hohenadler and S. Sorella.  This work was initiated at the KITP  during the workshop on  "Frustrated Magnetism and Quantum Spin Liquids From Theory and Models to Experiments" coordinated by  K. Kanoda, P. Lee, A.  Vishwanath and S. White and finalized at the MPI-PKS  (Dresden).  Funding from the DFG under the grant number AS 120/9-1 and  AS120/10-1 (Forschergruppe FOR 1807) is acknowledged. IFH has been suported by the NSERC of Canada.
We thank the  J\"ulich Supercomputing Centre  and the Leibniz-Rechenzentrum in M\"unich for generous allocation of CPU time.

\section{ Appendix}

Strictly speaking, the finite-size scaling form for the magnetization in presence of the external field is
\begin{equation}
m = L^{-\beta/\nu}   G ( \frac{L}{\xi}, h \xi^y )
= L^{-\beta/\nu}   G ( \frac{L}{\xi}, \frac{h_0}{L^d} \xi ^{y} ),
\end{equation}
where $G(X,Y)$ is the scaling function of {\it two} variables, and $\xi$ is the (diverging) correlation length.  Here we have  used the fact that the  relevant Fourier component of the local pinning field  scales as $\frac{h_0}{L^d}$.
The dimension $y$ of  the uniform external field is \cite{HerbutBook}
\begin{equation}
y= \frac{d+z+2-\eta}{2}.
\end{equation}
Away from criticality $\xi$ is  bounded, and  in the  thermodynamic limit,  the magnetization scales to  its field independent value.
The data in Fig. 1 confirms this and shows that  the  extrapolated value of the magnetization is $h_0$ independent provided that, for the considered lattice sizes,  $h_0$ is chosen to be large enough.  It is also interesting to point out that for values of $h_0$ in the range $ 1t < h_0 < 5 t $ the finite size value of the magnetization is next to independent on the value of the pinning field.

At criticality we can replace $\xi$ by $L$ to obtain:
\begin{equation}
m = L^{-\beta/\nu}   G ( 1,  h_0 L^{y-d} ).
\end{equation}
With the Lorentz symmetry at the critical point the value of the dynamical critical exponent is pinned to
$z=1$, and the scaling dimension of the {\it local field} $h_0$  is
\begin{equation}
y-d= \frac{3-d-\eta}{2}.
\end{equation}
In the $\epsilon$-expansion then,
\begin{equation}
y-d= \frac{\epsilon}{10}+ O(\epsilon^2),
\end{equation}
and rather small, presumably even for $\epsilon=1$ ($d=2$). For a fixed value of $h_0$, therefore, the second argument of the scaling function $G$ is almost constant.
To be more precise, we  can use the asymptotic form  $G( 1, Y)    \propto Y^{1/{\delta}} $  with $\frac{1}{\delta} =\frac{\beta}{\nu  y} $ \cite{HerbutBook} such that   in the  large-$L$ limit,   $m \propto h_0^{1/\delta}  L^{-\frac{\beta}{\nu}  +  \frac{(y-d)}{\delta} } $.  Within the $\epsilon$ expansion $\frac{(y-d)}{\delta} \simeq 0.05 $   which results in a  very small  correction for considered lattice sizes.

Hence on the whole,  the scaling function $G$ depends rather weakly on the second argument
and for practical purposes it suffices to neglect it.

\providecommand{\href}[2]{#2}\begingroup\raggedright\endgroup

\begin{thebibliography}{10}

\bibitem{Meng10}
Z.~Y. Meng, T.~C. Lang, S.~Wessel, F.~F. Assaad, and A.~Muramatsu, ``Quantum
  spin-liquid emerging in two-dimensional correlated Dirac fermions,'' {\em
  Nature} {\bfseries 464} (2010) 847--851.

\bibitem{Sorella12}
S.~Sorella, Y.~Otsuka, and S.~Yunoki, ``Absence of a Spin Liquid Phase in the
  Hubbard Model on the Honeycomb Lattice,''
  \href{http://dx.doi.org/http://dx.doi.org/10.1038/srep00992}{{\em Sci. Rep.}
  {\bfseries 2} (2012) 992}.

\bibitem{Sorella92}
S.~Sorella and E.~Tosatti, ``Semi-metal-insulator transition of the Hubbard
  model in the honeycomb lattice,'' {\em Europhys. Lett.} {\bfseries 19} (1992)
  699.

\bibitem{Paiva05}
T.~Paiva, R.~T. Scalettar, W.~Zheng, R.~R.~P. Singh, and J.~Oitmaa,
  ``Ground-state and finite-temperature signatures of quantum phase transitions
  in the half-filled Hubbard model on a honeycomb lattice,'' {\em Phys. Rev. B}
  {\bfseries 72} (2005) 085123.

\bibitem{Herbut06}
I.~F. Herbut, ``Interactions and phase transitions on graphene's honeycomb
  lattice,'' \href{http://dx.doi.org/10.1103/PhysRevLett.97.146401}{{\em Phys.
  Rev. Lett.} {\bfseries 97} (October, 2006) 146401}.
  \url{http://link.aps.org/doi/10.1103/PhysRevLett.97.146401}.

\bibitem{Herbut09a}
I.~F. Herbut, V.~Juri\ifmmode \check{c}\else \v{c}\fi{}i\ifmmode~\acute{c}\else
  \'{c}\fi{}, and O.~Vafek, ``Relativistic Mott criticality in graphene,''
  \href{http://dx.doi.org/10.1103/PhysRevB.80.075432}{{\em Phys. Rev. B}
  {\bfseries 80} (Aug, 2009) 075432}.
  \url{http://link.aps.org/doi/10.1103/PhysRevB.80.075432}.

\bibitem{Schmidt12}
H.-Y. Yang, A.~F. Albuquerque, S.~Capponi, A.~M. L\"auchli, and K.~P. Schmidt,
  ``Effective spin couplings in the Mott insulator of the honeycomb lattice
  Hubbard model,'' {\em New Journal of Physics} {\bfseries 14} no.~11, (2012)
  115027. \url{http://stacks.iop.org/1367-2630/14/i=11/a=115027}.

\bibitem{Chen2012}
C.-C. Chang and R.~T. Scalettar, ``Quantum Disordered Phase near the Mott
  Transition in the Staggered-Flux Hubbard Model on a Square Lattice,''
  \href{http://dx.doi.org/10.1103/PhysRevLett.109.026404}{{\em Phys. Rev.
  Lett.} {\bfseries 109} (Jul, 2012) 026404}.
  \url{http://link.aps.org/doi/10.1103/PhysRevLett.109.026404}.

\bibitem{Clark2013}
B.~K. Clark, ``Searching for Topological Degeneracy in the Hubbard Model with
  Quantum Monte Carlo,'' {\em arXiv:1305.0278} (2013) .

\bibitem{White07}
S.~R. White and A.~L. Chernyshev, ``N\'eel Order in Square and Triangular
  Lattice Heisenberg Models,''
  \href{http://dx.doi.org/10.1103/PhysRevLett.99.127004}{{\em Phys. Rev. Lett.}
  {\bfseries 99} (Sep, 2007) 127004}.
  \url{http://link.aps.org/doi/10.1103/PhysRevLett.99.127004}.

\bibitem{Hohenadler10}
M.~Hohenadler, T.~C. Lang, and F.~F. Assaad, ``Correlation Effects in Quantum
  Spin-Hall Insulators: A Quantum Monte Carlo Study,''
  \href{http://dx.doi.org/10.1103/PhysRevLett.106.100403}{{\em Phys. Rev.
  Lett.} {\bfseries 106} no.~10, (Mar, 2011) 100403}.
  \url{http://xxx.lanl.gov/abs/1011.5063}.

\bibitem{Hohenadler12}
M.~Hohenadler, Z.~Y. Meng, T.~C. Lang, S.~Wessel, A.~Muramatsu, and F.~F.
  Assaad, ``Quantum phase transitions in the Kane-Mele-Hubbard model,''
  \href{http://dx.doi.org/10.1103/PhysRevB.85.115132}{{\em Phys. Rev. B}
  {\bfseries 85} (Mar, 2012) 115132}.
  \url{http://link.aps.org/doi/10.1103/PhysRevB.85.115132}.

\bibitem{Rachel10}
S.~Rachel and K.~Le~Hur, ``Topological insulators and Mott physics from the
  Hubbard interaction,''
  \href{http://dx.doi.org/10.1103/PhysRevB.82.075106}{{\em Phys. Rev. B}
  {\bfseries 82} (Aug, 2010) 075106}.
  \url{http://link.aps.org/doi/10.1103/PhysRevB.82.075106}.

\bibitem{Assaad04}
F.~F. Assaad, ``Phase diagram of the half-filled two-dimensional SU(N)
  Hubbard-Heisenberg model: a quantum Monte Carlo study.,'' {\em Phys. Rev. B}
  {\bfseries 71} (2005) 075103.

\bibitem{Ulybyshev2013}
M.~V. Ulybyshev, P.~V. Buividovich, M.~I. Katsnelson, and M.~I. Polikarpov,
  ``Monte~Carlo Study of the Semimetal-Insulator Phase Transition in Monolayer
  Graphene with a Realistic Interelectron Interaction Potential,''
  \href{http://dx.doi.org/10.1103/PhysRevLett.111.056801}{{\em Phys. Rev.
  Lett.} {\bfseries 111} (Jul, 2013) 056801}.
  \url{http://link.aps.org/doi/10.1103/PhysRevLett.111.056801}.

\bibitem{Note1}
Clearly, one cannot exchange the limits in Eq. \ref {Lim.eq}. However, one can
  measure the magnetization at the largest distance available on the lattice,
  thereby tying together the thermodynamic and infinite distance from the
  pinning center limits. This procedure can lead to spurious results.

\bibitem{Assaad08_rev}
F.~F. {Assaad} and H.~G. {Evertz}, ``{World-line and Determinantal Quantum
  Monte Carlo Methods for Spins, Phonons and Electrons},'' in {\em
  Computational Many Particle Physics}, H.~{Fehske}, R.~{Schneider}, and
  A.~{Wei{\ss}e}, eds., vol.~739 of {\em Lecture Notes in Physics},
  pp.~277--356.
\newblock Springer Verlag, Berlin, 2008.

\bibitem{Capponi00}
S.~Capponi and F.~F. Assaad, ``Spin and charge dynamics of the ferromagnetic
  and antiferromagnetic two-dimensional half-filled Kondo lattice model,'' {\em
  Phys. Rev. B} {\bfseries 63} (2001) 155114.

\bibitem{Note2}
Here large means comparable to the bandwidth. If the pinning field is much
  larger than the bandwidth, charge fluctuations on the pinning site will be
  blocked by an energy scale set by $h_0$. Thereby in the limit $h_0
  \rightarrow \infty $ the pinning site effectively drops out of the
  Hamiltonian and no symmetry breaking occurs. The very slight drop in the
  magnetization at $h_0 =5$ and on small lattices in Fig. \ref
  {fig:Comparison}(a) could be a precursor of this effect.

\bibitem{Note3}
S. Wessel private communication.

\bibitem{Note4}
In Ref. \cite {Meng10} a small but finite spin gap with maximal value at
  $U/t=4$ was reported. The spin gap is determined with the same method as the
  single particle gap, by fitting the tail of the imaginary time displaced
  spin-spin correlation functions to a single exponential. Enhancing the
  imaginary time range for the fit produces a slight decrease of the spin gap
  for the larger lattices sizes. This in turn renders the extrapolation to the
  thermodynamic limit inconclusive.

\bibitem{HerbutBook}
I.~Herbut, {\em A Modern Approach to Critical Phenomena}.
\newblock Cambridge University Press, Cambridge, 2007.

\bibitem{Herbut09}
I.~F. Herbut, V.~Juri\ifmmode \check{c}\else \v{c}\fi{}i\ifmmode~\acute{c}\else
  \'{c}\fi{}, and B.~Roy, ``Theory of interacting electrons on the honeycomb
  lattice,'' \href{http://dx.doi.org/10.1103/PhysRevB.79.085116}{{\em Phys.
  Rev. B} {\bfseries 79} (Feb, 2009) 085116}.
  \url{http://link.aps.org/doi/10.1103/PhysRevB.79.085116}.

\end{thebibliography}

\end{document}